\documentclass[aps,prb,onecolumn,superscriptaddress,12pt]{revtex4-2}
\usepackage{amsmath,amssymb,bm}
\usepackage{graphicx}
\usepackage{xcolor}
\usepackage{orcidlink}
\usepackage{float}
\usepackage[ruled]{algorithm2e}
\usepackage{hyperref}

\hypersetup{
  colorlinks=true,
  linkcolor=blue,
  citecolor=blue,
  urlcolor=blue
}

\newcommand{\Heff}{H_{\mathrm{eff}}}
\newcommand{\GR}{G^{R}}

\newcommand{\Rcoh}{\mathcal{R}_{\mathrm{coh}}}
\newcommand{\Rmix}{\mathcal{R}_{\mathrm{mix}}}
\newcommand{\dd}{\mathrm{d}}
\newcommand{\ii}{\mathrm{i}}
\newcommand{\Neff}{N_{\mathrm{eff}}}
\newcommand{\Sresp}{S_{\mathrm{resp}}}
\newcommand{\rev}[1]{{\color{black}#1}}

\begin{document}

\title{Spectral Mixing, Skin Localization, and Linear Optical Response in Dissipative Photonic Lattices}

\author{Chen-Huan Wu
\orcidlink{0000-0003-1020-5977} }
\email{chenhuanwu1@gmail.com}
\affiliation{Department of Physics, Faculty of Science, Universiti Malaya, Kuala Lumpur 50603, Malaysia}



\begin{abstract}
\rev{We study the linear optical response of a finite dissipative Hatano--Nelson photonic lattice. The response between selected input and output ports is resolved into a phase-coherent intensity and an incoherent modal-weight contribution using the biorthogonal Green function. Their comparison isolates interference among non-Hermitian modal residues, while a response-weight entropy and the associated participation number $\Neff$ quantify how broadly the measured signal is distributed over the complex modes. The numerical results show that loss broadens the modal distribution, periodic-boundary spectral winding increases modal participation, and onsite disorder reduces it. Time-domain quantum-walk dynamics independently display the drift and right-edge accumulation produced by non-reciprocal hopping under open boundaries. A parameter map in the $(g,W)$ plane, supplemented by disorder-ensemble averages, identifies a finite-size crossover in which $\Neff$ responds to disorder before the response-weighted center is displaced from the skin boundary. The analysis applies to coupled waveguides, microring arrays, driven cavity lattices, and other linear photonic platforms with loss, gain, or non-reciprocal coupling.}
\end{abstract}

\maketitle

\section{Introduction}

Driven and dissipative photonic systems provide a natural setting for non-Hermitian physics. Optical loss, gain, radiative leakage, and non-reciprocal coupling can often be incorporated into an effective mode matrix whose eigenfrequencies are complex. This makes photonic lattices useful platforms for studying exceptional points, mode non-orthogonality, directional amplification, boundary-sensitive transport, and the non-Hermitian skin effect (NHSE)~\cite{ElGanainy2018,Miri2019,Parto2021}.

The Hatano--Nelson model is a minimal lattice model in which non-Hermiticity enters through an imaginary vector potential or, equivalently, asymmetric nearest-neighbor hopping~\cite{HatanoNelson1996,HatanoNelson1997}. In modern non-Hermitian band theory the same model is also the simplest setting for the NHSE: under open boundary conditions a macroscopic set of bulk modes acquires an exponential envelope and accumulates near one boundary, while under periodic boundary conditions the Bloch spectrum forms a loop in the complex plane~\cite{YaoWang2018,YaoSongWang2018,YokomizoMurakami2019}. This boundary sensitivity is not a secondary detail of the model. It is the main reason why response functions measured at a small number of ports can differ strongly from predictions based only on periodic-band eigenvalues.

\rev{Asymmetric hopping can also emerge from engineered dissipation rather than being imposed only at the Hamiltonian level. Biased dissipative couplings have been shown to generate directional motion and boundary accumulation in an open many-body lattice, while correlated gain and loss channels provide a second route to effective non-reciprocal transport~\cite{HuWangLianWang2025,GangulyAgarwalla2026}. These mechanisms motivate the effective parameter $g$ used below, although the present calculation remains a single-particle linear photonic model.}

In many experiments the accessible observables are not the full eigenvectors of the non-Hermitian operator. Instead, one measures transmission, reflection, local intensity, or a response spectrum between selected input and output ports. A useful theoretical description should therefore connect the global modal structure of a dissipative lattice to response functions built from a small number of local probes. This is especially important in a skin-effect system, because the input--output residues are exponentially biased by the side of the lattice on which the right and left eigenvectors accumulate.

\rev{The calculation proceeds in three stages. We first separate the coherent port-to-port intensity from the modal-weight response and use the resulting response entropy to count the modes visible to the probes. We then compare these frequency-domain quantities with a continuous-time photonic quantum walk, which resolves the drift and boundary accumulation of a localized excitation. Finally, the response is mapped in the $(g,W)$ plane and averaged over disorder realizations. The resulting finite-size crossover shows that disorder suppresses modal participation before it substantially shifts the response-weighted center away from the skin boundary. The coherent/incoherent separation follows standard wave-interference practice~\cite{Beenakker1997,AkkermansMontambaux2007}; its role here is to expose the biorthogonal residues of a finite non-Hermitian photonic chain.}


\section{Effective non-Hermitian photonic lattice}

We consider a finite set of coupled optical modes. In the single-excitation or classical linear-wave limit, the field amplitudes obey
\begin{equation}
  \ii \frac{\dd}{\dd t}|\psi(t)\rangle
  =
  \Heff |\psi(t)\rangle
  +
  |f_{\mathrm{in}}(t)\rangle ,
  \label{eq:eom}
\end{equation}
where $|f_{\mathrm{in}}(t)\rangle$ represents an external drive and
\begin{equation}
  \Heff
  =
  H_0 + V_{\mathrm{nr}} - \frac{\ii}{2}\Gamma .
  \label{eq:heff-general}
\end{equation}
Here $H_0$ is a Hermitian mode-coupling matrix, $V_{\mathrm{nr}}$ contains possible non-reciprocal or complex couplings, and $\Gamma$ describes local loss, radiative leakage, or effective gain--loss imbalance.

The concrete model used in the numerical examples is the dissipative Hatano--Nelson photonic chain
\begin{equation}
  \Heff
  =
  \sum_{n=1}^{L}
  \left(\omega_0+\Delta_n-\frac{\ii}{2}\kappa_n\right)
  |n\rangle\langle n|
  +
  \sum_{n=1}^{L-1}
  \left(
  J_R |n+1\rangle\langle n|
  +
  J_L |n\rangle\langle n+1|
  \right),
  \label{eq:photonic-chain}
\end{equation}
where the Hatano--Nelson parametrization is
\begin{equation}
  J_R=t e^{g},
  \qquad
  J_L=t e^{-g} .
  \label{eq:hn-hopping}
\end{equation}
$\Delta_n$ is a local detuning or onsite disorder drawn from a uniform random distribution $[-W/2, W/2]$ with $W$ being the disorder strength, and $\kappa_n$ is a local linewidth.
The parameter $g$ controls non-reciprocity, $\Delta_n$ is a local detuning or onsite disorder, and $\kappa_n$ is a local linewidth. Additional edge loss or inhomogeneous loss can be included by choosing site-dependent $\kappa_n$. The same notation also covers reciprocal dissipative lattices, coupled microrings, and finite waveguide arrays by choosing the appropriate matrix elements.
\rev{In an optical implementation, the asymmetry encoded by $g$ may arise from dynamic modulation, directional couplers, or engineered gain--loss channels. The last route is particularly relevant to dissipative platforms, where correlated channels can generate non-reciprocal propagation without introducing an explicit static imaginary gauge field~\cite{GangulyAgarwalla2026}.}

For the clean open chain with $\Delta_n=\kappa_n=0$, the asymmetric hopping in Eq.~\eqref{eq:photonic-chain} is related to the reciprocal chain by the non-unitary similarity transformation
\begin{equation}
  \Heff(g)=S(g)\Heff(0)S^{-1}(g),
  \qquad
  S_{nn}(g)=e^{gn} .
  \label{eq:similarity}
\end{equation}
Consequently, the open-chain eigenfrequencies are
\begin{equation}
  \Omega_m^{\mathrm{OBC}}
  =
  2t\cos\frac{m\pi}{L+1},
  \qquad
  m=1,\ldots,L,
  \label{eq:obc-spectrum}
\end{equation}
up to the common onsite frequency, whereas the right eigenmodes acquire the skin envelope
\begin{equation}
  \psi_m^R(n)
  \propto
  e^{gn}
  \sin\frac{m\pi n}{L+1} .
  \label{eq:right-skin-mode}
\end{equation}
For $g>0$ the right eigenvectors accumulate near the right boundary, while for $g<0$ they accumulate near the left boundary. The corresponding skin length is
\begin{equation}
  \xi_{\mathrm{skin}}
  \simeq
  \frac{1}{|g|}
  \label{eq:skin-length}
\end{equation}
when the lattice spacing is set to unity. This length scale provides a simple analytic reference for interpreting boundary response data. If the input and output ports are placed on opposite edges, the modal residue $\langle a|R_\nu\rangle\langle L_\nu|b\rangle$ contains the exponential boundary bias of the right and left eigenvectors. Therefore, the response participation number introduced below is sensitive not only to spectral overlap but also to skin-induced spatial accumulation.

Periodic boundary conditions remove the open-chain similarity reduction. The Bloch spectrum becomes
\begin{equation}
  \Omega^{\mathrm{PBC}}(k)
  =
  t e^{g}e^{-\ii k}+t e^{-g}e^{\ii k}
  =
  2t\cos k\cosh g
  -2\ii t\sin k\sinh g,
  \label{eq:pbc-loop}
\end{equation}
which forms a loop in the complex plane. The contrast between Eqs.~\eqref{eq:obc-spectrum} and \eqref{eq:pbc-loop} is the spectral counterpart of the NHSE.

\rev{The periodic spectral loop can be assigned a point-gap winding number around a reference energy $E_B$,}
\begin{equation}
  \rev{\nu_{\mathrm{pg}}(E_B)
  =
  \frac{1}{2\pi \ii}
  \int_{-\pi}^{\pi}\dd k\,
  \partial_k
  \ln\!\left[\Omega^{\mathrm{PBC}}(k)-E_B\right].}
  \label{eq:point-gap-winding}
\end{equation}
\rev{For $E_B=0$ inside the loop, the convention in Eq.~\eqref{eq:pbc-loop} gives $\nu_{\mathrm{pg}}=-1$ for $g>0$ and $\nu_{\mathrm{pg}}=+1$ for $g<0$. The sign reversal accompanies the reversal of the skin-accumulation edge under OBC. In multiband ladder models the winding can also depend on the reference-energy region, so different spectral sectors may accumulate at opposite boundaries~\cite{LiZeng2026}. The single-band chain studied here has one skin direction at fixed $g$, but Eq.~\eqref{eq:point-gap-winding} makes explicit that the PBC loop is a point-gap topological object rather than only a geometric spectral feature.}

The right and left eigenvectors are defined by
\begin{equation}
  \Heff |R_\nu\rangle = \Omega_\nu |R_\nu\rangle,
  \qquad
  \langle L_\nu|\Heff = \Omega_\nu \langle L_\nu|,
  \label{eq:biorthogonal-eigenproblem}
\end{equation}
with complex eigenfrequencies $\Omega_\nu$. We use the biorthogonal normalization
\begin{equation}
  \langle L_\mu|R_\nu\rangle=\delta_{\mu\nu},
  \qquad
  \sum_\nu |R_\nu\rangle\langle L_\nu|=\mathbb{I},
  \label{eq:biorthogonal-completeness}
\end{equation}
away from exceptional points. Close to exceptional points the spectral representation below should be replaced by its Jordan-chain form.

\begin{figure}[H]
    \centering
    \includegraphics[width=0.95\linewidth]{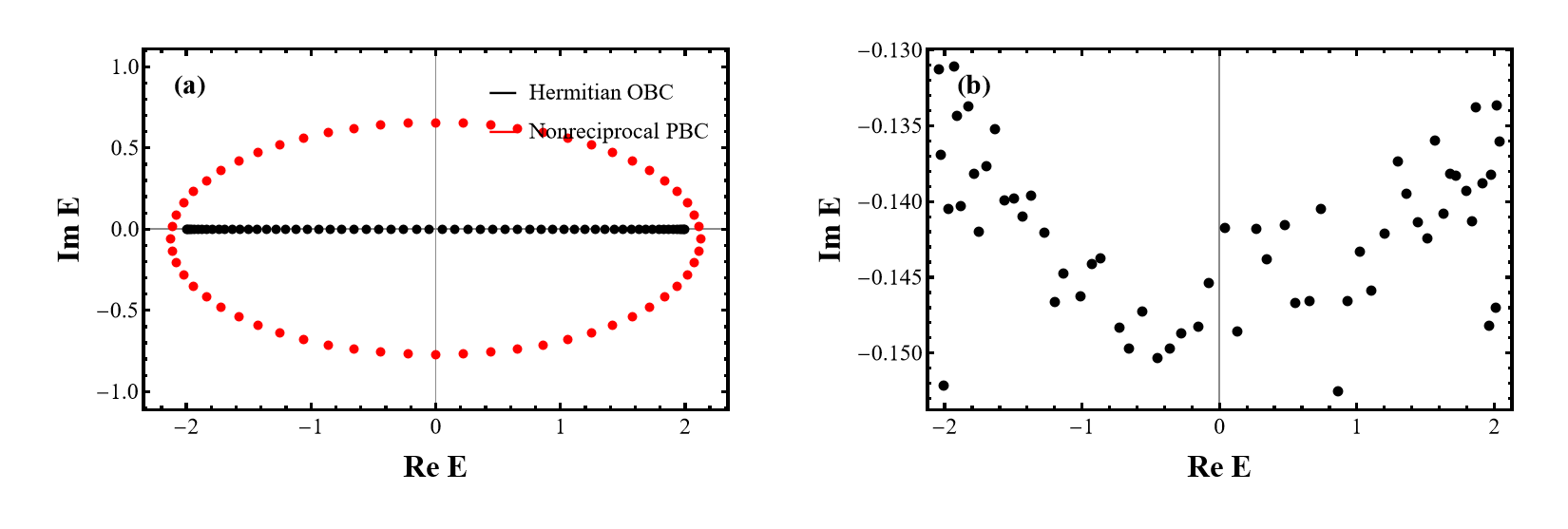}
    \caption{Complex spectra of finite one-dimensional photonic chains. (a) The black points show a Hermitian open-boundary chain, whose eigenfrequencies are real. The red points show a non-reciprocal periodic Hatano--Nelson chain, where the asymmetric hopping produces a complex spectral loop. The open-chain counterpart has skin-localized right eigenvectors with the envelope in Eq.~\eqref{eq:right-skin-mode}. (b) Spectrum of an open chain with non-reciprocal hopping, onsite disorder, uniform and inhomogeneous loss, and additional boundary loss. The eigenfrequencies acquire finite imaginary parts, corresponding to linewidths, and the real parts are irregularly spaced by disorder. All frequencies are measured in units of the hopping amplitude $t$.}
    \label{fig:complex-spectrum}
\end{figure}

Figure~\ref{fig:complex-spectrum} shows the corresponding finite-size spectra for representative chains. Panel (a) displays the standard spectral signature of the Hatano--Nelson NHSE: the open Hermitian reference has a real spectrum, while the non-reciprocal periodic chain produces a complex spectral loop. The associated open-chain right eigenvectors are not shown in the figure, but Eq.~\eqref{eq:right-skin-mode} gives their exponential skin profile. Panel (b) adds local loss, edge loss, onsite disorder, and non-reciprocal hopping in an open chain. The spectrum then occupies a finite-width complex band, with the imaginary part giving modal linewidths and the real part reflecting disordered resonance positions.

\section{Linear optical response}

For a monochromatic drive at frequency $\omega$, Eq.~\eqref{eq:eom} gives
\begin{equation}
  |\psi(\omega)\rangle
  =
  \GR(\omega)|f_{\mathrm{in}}(\omega)\rangle,
  \qquad
  \GR(\omega)
  =
  \frac{1}{\omega\mathbb{I}-\Heff} .
  \label{eq:retarded-green}
\end{equation}
A response from an input profile $|b\rangle$ to an output profile $|a\rangle$ is then
\begin{equation}
  \chi_{ab}(\omega)
  =
  \langle a|\GR(\omega)|b\rangle .
  \label{eq:chi-ab}
\end{equation}
In an input--output setting $\chi_{ab}$ may be converted into a transmission or reflection amplitude by including the external coupling rates. In the present work we focus only on the internal response kernel.

Using Eq.~\eqref{eq:biorthogonal-completeness}, the Green function has the spectral form
\begin{equation}
  \GR(\omega)
  =
  \sum_\nu
  \frac{|R_\nu\rangle\langle L_\nu|}
       {\omega-\Omega_\nu} .
  \label{eq:spectral-green}
\end{equation}
Thus
\begin{equation}
  \chi_{ab}(\omega)
  =
  \sum_\nu
  \frac{\langle a|R_\nu\rangle\langle L_\nu|b\rangle}
       {\omega-\Omega_\nu} .
  \label{eq:chi-spectral}
\end{equation}
Equation~\eqref{eq:chi-spectral} is the basic optical analogue of a Kubo-type spectral response formula. The complex poles determine the resonance positions (real part) and linewidths (imaginary part), while the residues encode how strongly each biorthogonal mode couples to the chosen input and output channels. In a skin-effect chain these residues depend strongly on whether the probes are placed near the skin-accumulation edge, the opposite edge, or the bulk.

For numerical work it is useful to introduce a finite broadening,
\begin{equation}
  \delta_\gamma(x)
  =
  \frac{1}{\pi}\frac{\gamma}{x^2+\gamma^2},
  \label{eq:lorentzian}
\end{equation}
and to use either the complex response $\chi_{ab}(\omega)$ or an absorptive spectral function such as
\begin{equation}
  A_{ab}(\omega)
  =
  -2\,\mathrm{Im}\,\chi_{ab}(\omega).
  \label{eq:absorptive}
\end{equation}
For a passive stable system the measured intensity is often closer to $|\chi_{ab}(\omega)|^2$ or to a properly normalized transmission coefficient, depending on the coupling geometry.

\section{Coherent and modal-weight response channels}

For a fixed input profile $|b\rangle$ and output profile $|a\rangle$, Eq.~\eqref{eq:chi-spectral} can be written as a sum of modal amplitudes,
\begin{equation}
  \chi_{ab}(\omega)
  =
  \sum_\nu A_\nu^{ab}(\omega),
  \qquad
  A_\nu^{ab}(\omega)
  =
  \frac{\langle a|R_\nu\rangle\langle L_\nu|b\rangle}
       {\omega+\ii\eta-\Omega_\nu} .
  \label{eq:modal-amplitude}
\end{equation}
Here $\eta$ is a small numerical broadening in addition to the intrinsic linewidth contained in $\Omega_\nu$. The intensity measured between the two ports is phase sensitive and contains interference between different biorthogonal modes. We define this coherent response as
\begin{equation}
  \Rcoh(\omega)
  =
  \left|
  \sum_\nu A_\nu^{ab}(\omega)
  \right|^2 .
  \label{eq:rcoh}
\end{equation}
To isolate the contribution of modal weights without retaining the relative phases between modes, we also define the modal-weight response
\begin{equation}
  \Rmix(\omega)
  =
  \sum_\nu |A_\nu^{ab}(\omega)|^2 .
  \label{eq:rmix}
\end{equation}
The two quantities are related by
\begin{equation}
  \Rcoh(\omega)
  =
  \Rmix(\omega)
  +
  \sum_{\mu\neq\nu}
  A_\mu^{ab}(\omega)
  \left[A_\nu^{ab}(\omega)\right]^* .
  \label{eq:interference-term}
\end{equation}
This decomposition is a direct coherent-versus-incoherent separation, analogous to familiar treatments of wave interference, mesoscopic conductance fluctuations, coherent backscattering, and speckle statistics~\cite{Beenakker1997,AkkermansMontambaux2007}. The useful point here is not the algebra itself, but the fact that in a non-Hermitian lattice the modal amplitudes $A_\nu^{ab}$ contain biorthogonal residues that are strongly reshaped by non-normality, loss, and skin localization. Thus $\Rcoh$ and $\Rmix$ can differ even when they are built from the same set of complex poles.

In a nearly Hermitian system with well-separated resonances, the two curves become similar near isolated peaks. In a lossy or strongly non-normal lattice, however, the incoherent modal weights may spread over a broader frequency interval even when the coherent response remains concentrated near a smaller number of constructive-interference resonances. In a Hatano--Nelson chain, the same comparison also depends on the probe geometry: a pair of boundary probes placed along or against the skin direction can emphasize different modal residues.

The input and output vectors in Eqs.~\eqref{eq:modal-amplitude}--\eqref{eq:rmix} may be single sites, boundary ports, or normalized vectors supported on a finite window. For instance, replacing $|a\rangle$ and $|b\rangle$ by window-supported profiles makes the same definitions applicable to local density probes and spatially averaged measurements.

\section{Response-weight entropy and skin diagnostics}

The modal-weight response in Eq.~\eqref{eq:rmix} gives a natural probability distribution over the modes contributing at a fixed frequency. We define
\begin{equation}
  w_\nu(\omega)
  =
  |A_\nu^{ab}(\omega)|^2,
  \qquad
  p_\nu(\omega)
  =
  \frac{w_\nu(\omega)}
       {\sum_\mu w_\mu(\omega)} .
  \label{eq:response-probability}
\end{equation}
The corresponding Shannon entropy is
\begin{equation}
  \Sresp(\omega)
  =
  -\sum_\nu p_\nu(\omega)\ln p_\nu(\omega),
  \label{eq:response-entropy}
\end{equation}
and the effective number of participating modes is
\begin{equation}
  \Neff(\omega)
  =
  \exp\left[\Sresp(\omega)\right] .
  \label{eq:neff}
\end{equation}
With this convention $\Neff\approx 1$ for a single dominant resonance and $\Neff\approx M$ when $M$ modes contribute with comparable weights. The inverse-participation definition $(\sum_\nu p_\nu^2)^{-1}$ gives a related second-order participation number and leads to the same qualitative diagnostics in the examples considered below.

This entropy should not be confused with a thermodynamic entropy unless the driven-dissipative system is coupled to a bath that enforces detailed balance. It is instead a spectral participation entropy: it measures how the measured response is distributed over the non-Hermitian mode basis. In a narrow isolated resonance $\Neff\approx 1$, whereas strongly overlapping resonances or dissipative mode mixing can give larger values.

For skin-effect systems the same quantity has a spatial interpretation through the residues $\langle a\vert{}R_\nu\rangle\langle L_\nu\vert{}b\rangle$ in Eq.~\eqref{eq:modal-amplitude}. A right-localized skin mode can have a large overlap with a probe on the skin side and an exponentially small overlap with a probe on the opposite side. For the clean chain, Eq.~\eqref{eq:right-skin-mode} gives the approximate envelope scale $\xi_{\mathrm{skin}}=1/|g|$. Thus a boundary-resolved version of Eq.~\eqref{eq:neff}, obtained by changing $|a\rangle$ and $|b\rangle$ between left, center, and right windows, can be used to detect whether the measured response is dominated by skin accumulation or by ordinary spectral broadening. This is the main reason that a response-level participation number is useful in the Hatano--Nelson photonic chain.

A simple additional diagnostic is the center of mass of the right eigenmode,
\begin{equation}
  X_\nu^R
  =
  \frac{\sum_{n=1}^L n |\langle n|R_\nu\rangle|^2}
       {\sum_{n=1}^L |\langle n|R_\nu\rangle|^2} .
  \label{eq:skin-com}
\end{equation}
\rev{The distribution $p_\nu(\omega)$ also gives a mode-resolved diagnostic of the spatial origin of the response. Instead of reducing the response to a single scalar $\Neff(\omega)$, one can weight any modal observable by $p_\nu(\omega)$. In particular, the response-weighted skin center is}
\begin{equation}
  \rev{X_{\mathrm{resp}}^R(\omega)
  =
  \sum_\nu p_\nu(\omega) X_\nu^R .}
  \label{eq:response-com}
\end{equation}
\rev{For a right-skin chain, $X_{\mathrm{resp}}^R(\omega)$ moves toward the right boundary when the measured signal is dominated by skin-localized modes. This response-weighted center is used below as the spatial counterpart of $\Neff(\omega)$.}
For the single-band chain, the response-weighted center $X_{\text{resp}}^R$ provides a smooth directional measure of skin accumulation under frequency averaging and weak disorder.

\begin{figure}[H]
    \centering
    \includegraphics[width=1\linewidth]{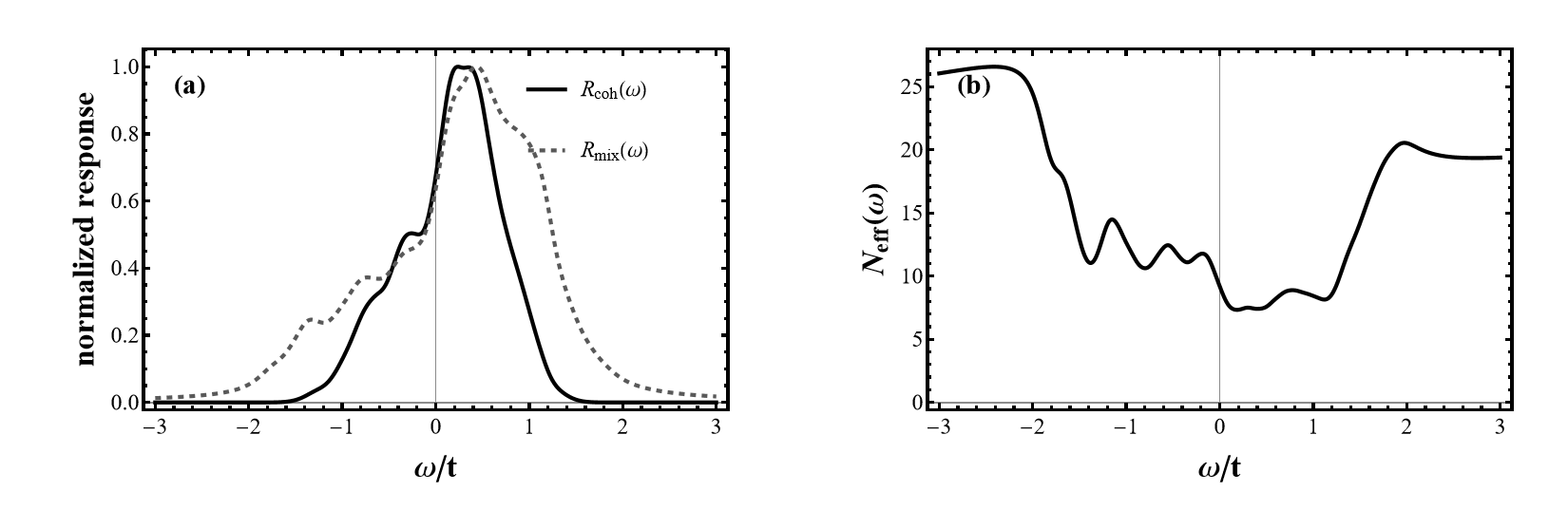}
    \caption{Boundary-to-boundary optical response and modal participation in a finite lossy Hatano--Nelson photonic chain. (a) Normalized coherent response $\Rcoh(\omega)$ and modal-weight response $\Rmix(\omega)$. The coherent response retains phase interference between biorthogonal modal amplitudes, whereas $\Rmix$ keeps only the incoherent modal weights. The broader dashed curve indicates that dissipative spectral mixing distributes modal weight over a wider frequency range than the coherent output intensity. (b) Effective number of participating modes $\Neff(\omega)=\exp[\Sresp(\omega)]$. Because the residues contain the boundary overlaps $\langle a|R_\nu\rangle\langle L_\nu|b\rangle$, the same diagnostic can be used to test skin-biased boundary accumulation by moving the input and output windows. Frequencies are measured in units of the hopping amplitude $t$.}
    \label{fig:response-neff}
\end{figure}

For $g>0$ in the clean open chain, $X_\nu^R$ shifts toward the right boundary with a length scale controlled by $\xi_{\mathrm{skin}}$. In disordered chains the Anderson localization length competes with this skin length. For weak onsite disorder with a uniform distribution of width $W$, the one-dimensional Anderson localization length near the band center scales as
\begin{equation}
  \xi_{\mathrm{A}}(E\simeq 0)
  \sim
  \frac{96t^2}{W^2},
  \label{eq:anderson-length}
\end{equation}
up to model-dependent numerical factors. The response is expected to be skin dominated when $\xi_{\mathrm{skin}}\ll \xi_{\mathrm{A}}$ and disorder dominated when the reverse inequality holds. Equation~\eqref{eq:anderson-length} is only an order-of-magnitude estimate, but it gives an analytic scale against which the numerical $\Neff$ trends can be compared.
Figure~\ref{fig:response-neff} gives a representative boundary-to-boundary calculation for a finite lossy and disordered chain. The coherent response is sharply concentrated around its main constructive-interference peak, while the modal-weight response is broader and has a more extended high-frequency shoulder. This shows that the same non-Hermitian mode set can produce a narrow measured coherent signal even when the modal weights are spread over a larger frequency interval. The participation number in Fig.~\ref{fig:response-neff}(b) is largest where many weak modal residues overlap and is reduced near the central frequency region, where the response is concentrated into fewer effective modes. The minimum of $\Neff(\omega)$ does not necessarily coincide with the largest response intensity, because it is controlled by the distribution of modal weights rather than by the absolute signal strength. In a skin-effect geometry, changing the input and output ports from one edge to the other would change this distribution through the exponential factors in Eq.~\eqref{eq:right-skin-mode}.

\section{Numerical simulation}

The formulation above can be tested in small finite lattices without large computational resources. A typical calculation proceeds as follows. First, choose a finite matrix $\Heff$ such as Eq.~\eqref{eq:photonic-chain}. Disorder may be added through $\Delta_n$, local loss through $\kappa_n$, and non-reciprocity through $J_R\neq J_L$. Second, compute the left and right eigenvectors and impose the biorthogonal normalization in Eq.~\eqref{eq:biorthogonal-completeness}. Third, evaluate the local response functions $\chi_{ab}(\omega)$, $\Rcoh(\omega)$, $\Rmix(\omega)$, and $\Sresp(\omega)$ on a frequency grid. Finally, compare bulk and boundary probes. For example, one may choose
\begin{equation}
  D_{\mathrm{L}}=\sum_{n=1}^{w}|n\rangle\langle n|,
  \qquad
  D_{\mathrm{C}}=\sum_{n=L/2-w/2}^{L/2+w/2}|n\rangle\langle n|,
  \qquad
  D_{\mathrm{R}}=\sum_{n=L-w+1}^{L}|n\rangle\langle n| .
  \label{eq:windows}
\end{equation}
The comparison between $D_{\mathrm{L/R}}$ and $D_{\mathrm{C}}$ separates boundary-sensitive skin accumulation from bulk spectral broadening or disorder-induced localization.

The boundary-to-boundary response is
\begin{equation}
  T_{\mathrm{LR}}(\omega)
  =
  |\langle L|\GR(\omega)|R\rangle|^2,
  \label{eq:tlr}
\end{equation}
where $|L\rangle$ and $|R\rangle$ denote left and right input profiles.
The local spectral density is
\begin{equation}
  \rho_A(\omega)
  =
  -\frac{1}{\pi}\mathrm{Im}\,
  \mathrm{Tr}\left[
  D_A \GR(\omega)
  \right].
  \label{eq:local-density}
\end{equation}
Together with $\Sresp(\omega)$ and the eigenmode center of mass in Eq.~\eqref{eq:skin-com}, these quantities give a small set of diagnostics for finite dissipative photonic lattices.

\rev{The same finite matrix can also be used as a time-domain continuous-time photonic quantum walk generator. Starting from a localized single-photon or classical field excitation at site $n_0$,}
\begin{equation}
  \rev{|\psi(t)\rangle
  =
  e^{-\ii \Heff t}|n_0\rangle .}
  \label{eq:time-domain-state}
\end{equation}
\rev{Because $\Heff$ is non-Hermitian, the norm is generally not conserved. It is therefore useful to work with the normalized intensity distribution}
\begin{equation}
  \rev{P_n(t)
  =
  \frac{|\langle n|\psi(t)\rangle|^2}
       {\sum_m |\langle m|\psi(t)\rangle|^2}.}
  \label{eq:normalized-intensity}
\end{equation}
\rev{The corresponding wave-packet center and width are}
\begin{equation}
  \rev{\bar n(t)=\sum_n nP_n(t),
  \qquad
  \sigma(t)=
  \left[\sum_n (n-\bar n(t))^2P_n(t)\right]^{1/2}.}
  \label{eq:walk-width}
\end{equation}
\rev{These are the single-particle analogues of standard transport diagnostics used in continuous-time quantum walks~\cite{ForghieriParis2026}. To track the NHSE directly, we use the normalized weight in a right-edge window,}
\begin{equation}
  \rev{P_{\mathrm{skin}}(t;\ell)=\sum_{n=L-\ell+1}^{L}P_n(t) .}
  \label{eq:walk-return-skin}
\end{equation}
\rev{The pair $\sigma(t)$ and $P_{\mathrm{skin}}(t;\ell)$ distinguishes spatial spreading from directed accumulation at the skin edge.}

\begin{figure}[H]
    \centering
    \includegraphics[width=1\linewidth]{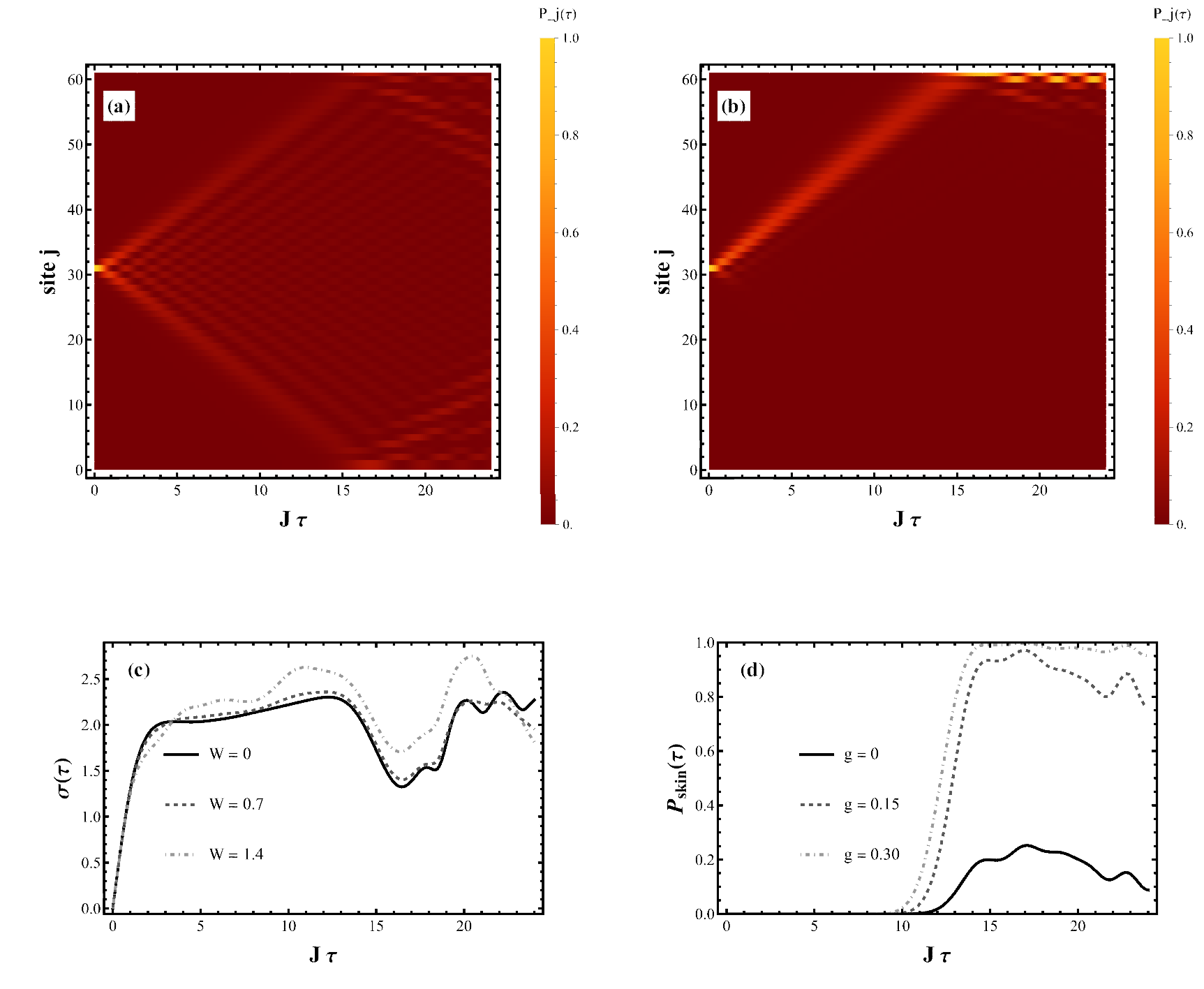}
    \caption{\rev{Time-domain non-Hermitian quantum-walk diagnostics for the dissipative Hatano--Nelson photonic chain. (a) Normalized intensity distribution $P_j(\tau)$ for a reciprocal open chain with $g=0$, initialized at the central site. The two counter-propagating branches show approximately symmetric spreading before finite-size boundary reflections become visible. (b) The same quantity for a non-reciprocal chain with $g=0.30$. The intensity is biased toward the right boundary and forms a long-lived edge-accumulated branch, as expected from the skin envelope of the right eigenmodes. (c) Wave-packet width $\sigma(\tau)$ for several onsite-disorder strengths. For the selected finite chain the width grows rapidly at short times and then exhibits boundary-reflection and disorder-dependent oscillations; it should therefore be viewed as a finite-time spreading diagnostic rather than a clean asymptotic localization exponent. (d) Right-edge skin probability $P_{\mathrm{skin}}(\tau)$, evaluated over the last six sites, for different non-reciprocity parameters. The reciprocal chain shows only a weak edge weight caused by finite-size reflections, whereas nonzero $g$ drives a rapid increase of boundary occupation toward values close to unity. Time is measured in units of $J^{-1}$.}}
    \label{fig:time-domain-walk}
\end{figure}

\rev{Figure~\ref{fig:time-domain-walk} applies these time-domain diagnostics to an open chain initialized at the central site. The density plots in Figs.~\ref{fig:time-domain-walk}(a) and \ref{fig:time-domain-walk}(b) use the normalized intensity in Eq.~\eqref{eq:normalized-intensity}, so the color scale measures spatial redistribution of the surviving field rather than the total decay. In the reciprocal case, the excitation spreads in both directions and later reflects from the finite boundaries. With non-reciprocal hopping, the dominant intensity branch is displaced toward increasing site index and accumulates near the right edge, giving a direct time-domain manifestation of the NHSE.}

\rev{In Fig.~\ref{fig:time-domain-walk}(c), the width $\sigma(\tau)$ remains a useful measure of spatial spreading, but in a non-Hermitian skin chain it is not by itself a monotonic measure of transport distance: after the packet reaches the boundary, the normalized distribution can remain narrow even though it has moved far from the initial site. The boundary weight $P_{\mathrm{skin}}(\tau)$ is therefore the more direct NHSE diagnostic. Its strong growth for $g=0.15$ and $g=0.30$ confirms that the right-boundary accumulation inferred from the spectral skin profile also appears in real-time propagation. This time-domain result complements the frequency-domain participation number: $\Neff(\omega)$ resolves how many biorthogonal poles contribute to a driven response, while $P_{\mathrm{skin}}(\tau)$ shows where the corresponding non-unitary propagation deposits the field intensity.}
\rev{The boundary is essential to this distinction. Dissipative many-body models likewise show directional motion under periodic boundaries but persistent density accumulation only when an open edge is present~\cite{HuWangLianWang2025}. The present single-particle result is the corresponding photonic limit: non-reciprocity sets the drift, while the open boundary converts it into a skin profile.}

\begin{figure}[H]
    \centering
    \includegraphics[width=1\linewidth]{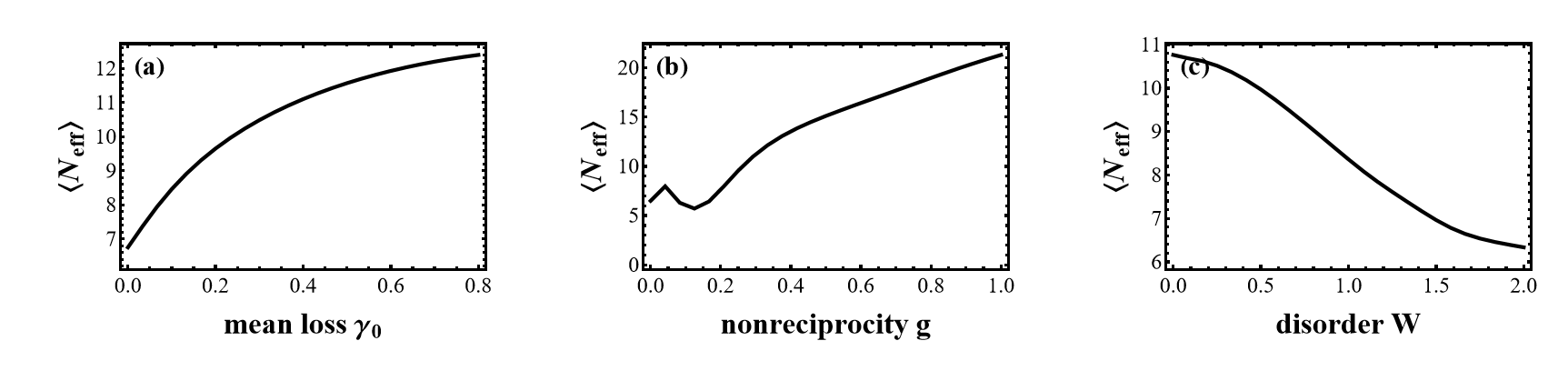}
    \caption{Frequency-averaged response participation number as a function of the main non-Hermitian lattice parameters. The chain length is $L=28$ and all frequencies are measured in units of $t$. (a) Increasing the mean loss $\gamma_0$ broadens the resonances and increases $\langle\Neff\rangle$, indicating that more modes contribute to the measured response. (b) Increasing the non-reciprocity parameter $g$ enhances modal participation after a small weak-$g$ dip. This sweep is evaluated in a periodic chain so that the non-reciprocal spectral winding is not removed by the open-chain similarity transformation. The small dip is not interpreted as a separate phase or transition. (c) Increasing disorder $W$ reduces $\langle\Neff\rangle$, consistent with spatial and spectral localization suppressing the number of modes visible to the boundary response. The displayed disorder sweep is a representative finite-size realization rather than a many-realization disorder average.}
    \label{fig:neff-sweeps}
\end{figure}

In the numerical figures below the chain length is $L=28$ for the parameter sweeps and $L=44$ for the frequency-resolved response. \rev{The time-domain quantum-walk calculation in Fig.~\ref{fig:time-domain-walk} uses $L=61$ and a central initial site.} The frequency grid is finite and all frequencies are measured in units of $t$. The curves in Fig.~\ref{fig:neff-sweeps} use a fixed representative disorder realization for each sweep rather than a disorder ensemble average. Therefore, the curves should be read as finite-size diagnostic trends. For quantitative disorder statistics one should average $\Neff$ over many independent disorder realizations and quote the standard error.

We now use the response participation number to compare the effects of loss, non-reciprocity, and disorder. The plotted quantity is the frequency-averaged participation number,
\begin{equation}
  \langle \Neff\rangle
  =
  \frac{1}{\omega_2-\omega_1}
  \int_{\omega_1}^{\omega_2}
  \Neff(\omega)\,\dd\omega,
  \label{eq:averaged-neff}
\end{equation}
evaluated over a fixed finite frequency window. The average is not a thermodynamic quantity; it is a compact summary of how broadly the measured response is distributed over the non-Hermitian modes.

The loss sweep in Fig.~\ref{fig:neff-sweeps}(a) shows a monotonic increase of $\langle\Neff\rangle$. This is the expected behavior for a passive finite chain in the chosen parameter regime. Larger local linewidths broaden nearby resonances and increase their spectral overlap, so a fixed input--output response samples a larger set of modes. The curve gradually saturates because once the linewidths exceed the typical level spacing, further loss changes the relative modal weights less efficiently than at small loss.

The non-reciprocity sweep in Fig.~\ref{fig:neff-sweeps}(b) is evaluated with periodic boundary conditions. This choice is deliberate. In an open Hatano--Nelson chain, asymmetric hopping can be removed from the spectrum by the non-unitary transformation in Eq.~\eqref{eq:similarity}, although the corresponding right eigenvectors remain skin localized. Periodic boundary conditions prevent this cancellation and expose the complex spectral winding in Eq.~\eqref{eq:pbc-loop}. The resulting curve shows a small dip at weak non-reciprocity followed by a pronounced increase at larger $g$. The weak-$g$ dip is a finite-size and finite-broadening feature of the selected parameter set; no phase boundary is assigned to it. The robust trend is the large-$g$ growth, which reflects stronger spectral winding and increased overlap among the modal residues seen by the probes.

\begin{figure}[H]
    \centering
    \includegraphics[width=1\linewidth]{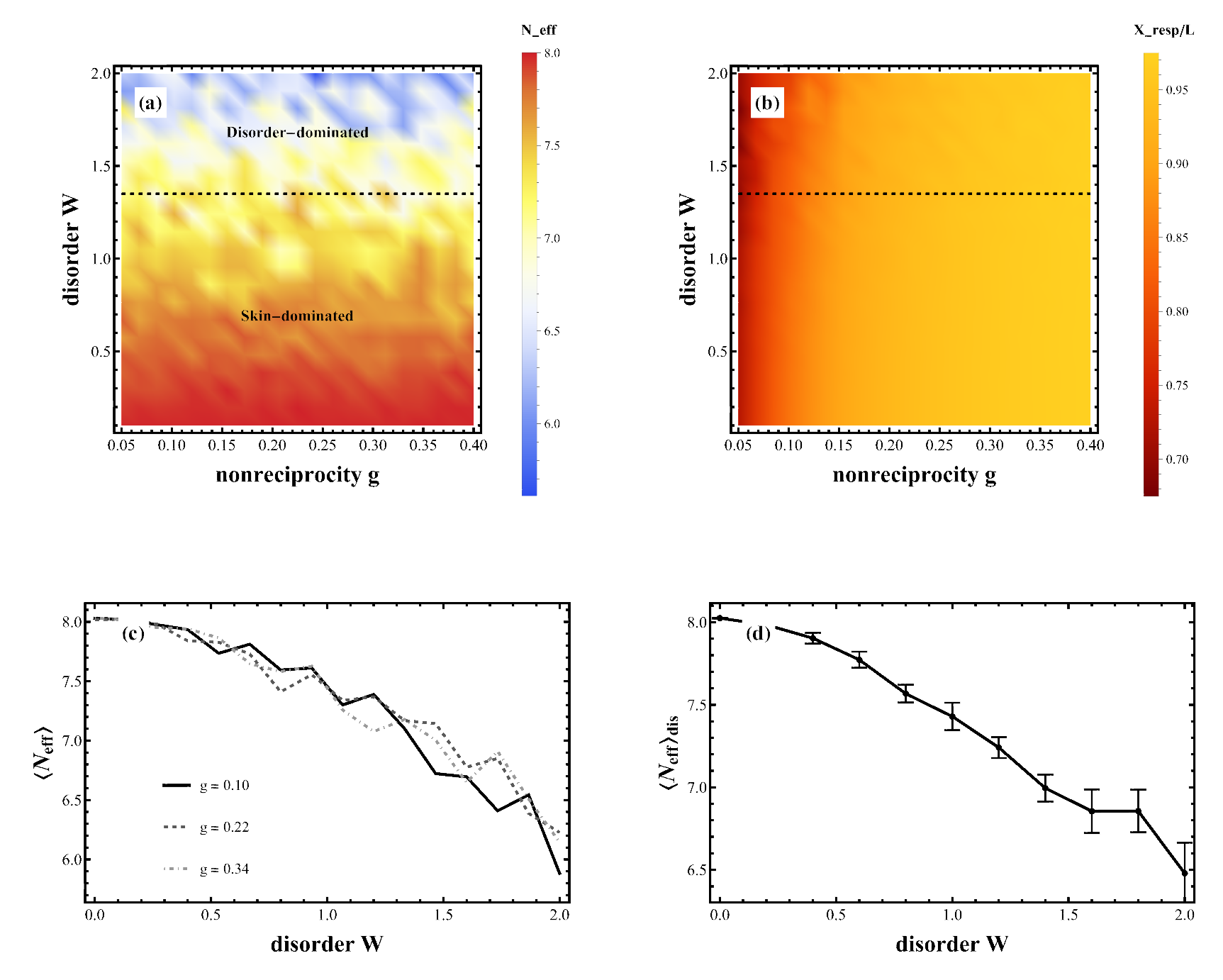}
    \caption{\rev{Finite-size validation of the skin--disorder response crossover. (a) Frequency-averaged response participation number $\langle\Neff\rangle$ in the $(g,W)$ plane for an open dissipative Hatano--Nelson chain of length $L=26$. The data are averaged over a finite frequency window and over five disorder realizations at each grid point. The dashed horizontal line marks the empirical finite-size crossover scale $W_{\mathrm{cross}}\simeq1.35$ separating a skin-dominated response regime from a disorder-suppressed response regime. (b) Response-weighted right-eigenmode center $\langle X_{\mathrm{resp}}^R\rangle/L$ over the same parameter grid. In contrast to $\langle\Neff\rangle$, the spatial center remains close to the right boundary throughout most of the scanned region, showing that the skin bias is robust even when disorder already suppresses modal participation. (c) Disorder dependence of $\langle\Neff\rangle$ for three fixed non-reciprocity values, averaged over eight disorder realizations for each point. (d) Disorder ensemble average $\langle\Neff\rangle_{\mathrm{dis}}$ at $g=0.22$ with standard-error bars computed from fifteen disorder realizations. The monotonic decrease confirms that the suppression of modal participation by disorder is not a single-realization artifact.}}
    \label{fig:skin-disorder-validation}
\end{figure}

The disorder sweep in Fig.~\ref{fig:neff-sweeps}(c) has the opposite trend. Increasing $W$ reduces the averaged participation number, meaning that fewer modes dominate the response within the same frequency window. Physically, onsite disorder separates and localizes the mode profiles, reducing their simultaneous overlap with the chosen input and output channels. This behavior is consistent with the estimate in Eq.~\eqref{eq:anderson-length}: stronger disorder shortens $\xi_{\mathrm{A}}$ and eventually competes with or dominates the skin length $\xi_{\mathrm{skin}}$. 
The ensemble calculation in Fig.~\ref{fig:skin-disorder-validation} is used below to separate this systematic trend from sample-to-sample fluctuations. 

\rev{Figure~\ref{fig:skin-disorder-validation} tests the length-scale interpretation in Eqs.~\eqref{eq:skin-length} and \eqref{eq:anderson-length} more directly than the one-parameter sweeps. In the open chain, once $g>0$ is finite, the right eigenvectors acquire the exponential skin envelope in Eq.~\eqref{eq:right-skin-mode}. The heat map in Fig.~\ref{fig:skin-disorder-validation}(a) shows that the strongest variation of $\langle\Neff\rangle$ then occurs along the disorder direction: increasing $W$ reduces the number of biorthogonal modes that simultaneously contribute to the boundary response. The nearly horizontal color bands mean that, for the finite sizes considered here, onsite disorder controls the spectral participation more strongly than moderate changes in $g$.}

\rev{The dashed line in Fig.~\ref{fig:skin-disorder-validation}(a,b) should be understood as a finite-size crossover scale rather than a thermodynamic critical boundary. Equating the asymptotic estimates $\xi_{\mathrm{skin}}=1/|g|$ and $\xi_{\mathrm{A}}\sim96t^2/W^2$ would give the parametric curve}
\begin{equation}
  \rev{W\simeq \sqrt{96t^2|g|}.}
  \label{eq:asymptotic-skin-anderson-boundary}
\end{equation}
\rev{However, this estimate is derived for an infinite one-dimensional disordered bulk. In the present finite photonic chain with $L=26$ and $t=1$, the empirical crossover around $W_{\mathrm{cross}}\simeq1.35$ still corresponds to}
\begin{equation}
  \rev{\xi_{\mathrm{A}}(W_{\mathrm{cross}})
  \sim
  \frac{96}{1.35^2}
  \simeq
  5.3\times10^1
  > L .}
  \label{eq:finite-size-xiA-estimate}
\end{equation}
\rev{Thus the numerical data do not represent a sharp Anderson transition. They show a finite-size crossover in which disorder-induced dephasing and mode separation begin to suppress the response participation even before a strict bulk localization length becomes shorter than the simulated chain. This is why the visually useful guide in Fig.~\ref{fig:skin-disorder-validation} is drawn as the empirical horizontal scale $W_{\mathrm{cross}}\simeq1.35$, while Eq.~\eqref{eq:asymptotic-skin-anderson-boundary} is kept as an asymptotic reference.}

\rev{Panel (b) demonstrates that $\langle X_{\mathrm{resp}}^R\rangle/L$ is less sensitive to the same moderate disorder. This difference is physically useful. The response participation number probes the distribution of modal residues in frequency space, so it decreases when disorder destroys phase coherence and separates the modal weights. The spatial center instead measures the average boundary bias of the right eigenvectors. For a clean skin envelope $|\psi_n^R|^2\propto e^{2gn}$, a continuum estimate gives}
\begin{equation}
  \rev{\frac{\bar n}{L}
  \simeq
  1-\frac{1}{2gL}}
  \label{eq:skin-com-estimate}
\end{equation}
\rev{when $gL$ is sufficiently large. Hence even a moderate non-reciprocity can pin the response-weighted center close to the right boundary. For example, at $g=0.22$ and $L=26$, Eq.~\eqref{eq:skin-com-estimate} gives $\bar n/L\simeq0.91$, consistent with the high values in Fig.~\ref{fig:skin-disorder-validation}(b). The important conclusion is that $X_{\mathrm{resp}}^R$ confirms the persistence of skin-biased spatial localization, whereas $\Neff$ is a more sensitive probe of disorder-induced modal decoherence.}

\rev{The lower panels provide two checks on this interpretation. Figure~\ref{fig:skin-disorder-validation}(c) shows that the decay of $\langle\Neff\rangle$ with $W$ persists for several representative values of $g$, confirming that the disorder trend is not tied to a single non-reciprocity choice. Figure~\ref{fig:skin-disorder-validation}(d) repeats the disorder sweep at $g=0.22$ with an explicit disorder ensemble and standard-error bars. The averaged curve keeps the same monotonic suppression observed in the representative sweep, so the decrease of response participation with disorder is not an artifact of one random potential.}

\rev{More generally, non-Hermitian critical scaling can involve a skin-depth scale in addition to a conventional bulk correlation or localization length, so scaling arguments based on a single diverging length need not remain valid under open boundary conditions~\cite{AroucaLeeMoraisSmith2020}. The present calculation does not establish a non-Bloch critical point or extract critical exponents. It instead supports the more limited conclusion that $W_{\mathrm{cross}}\simeq1.35$ is a finite-size optical-response crossover, rather than a universal thermodynamic phase boundary.}

Thus Figs.~\ref{fig:response-neff} and \ref{fig:neff-sweeps} support the use of $\Neff$ as a response-level diagnostic. Loss increases modal overlap, disorder suppresses modal participation, and non-reciprocity becomes visible either through open-chain skin accumulation in the eigenvectors or through periodic spectral winding in the complex spectrum. \rev{The time-domain calculation in Fig.~\ref{fig:time-domain-walk} adds the complementary real-time picture: a localized input evolves into a boundary-accumulated wave packet when non-reciprocity is present.} 

\section{Discussion}

The main point of this paper is not that local or diagonal probes are always sufficient. Rather, the question is when a restricted optical response already contains enough information to distinguish coherent intermode response from dissipative spectral mixing and skin-biased boundary accumulation. In finite non-Hermitian photonic systems this distinction is experimentally meaningful because one often controls only a few ports or local measurement windows.

The coherent channel in Eq.~\eqref{eq:rcoh} retains off-diagonal interference between modal amplitudes and is sensitive to phase coherence between modes. The modal-weight channel in Eq.~\eqref{eq:rmix} removes this information and keeps only the diagonal weights in the mode index. Their difference therefore diagnoses the extent to which interference between non-Hermitian modal residues affects the measured response. The response entropy in Eq.~\eqref{eq:response-entropy} then quantifies whether the response is carried by a few resonances or by a broad set of mixed modes.

\rev{Eigenvector nonorthogonality enters this comparison separately from ordinary linewidth broadening. In critical non-Hermitian free-fermion steady states, overlap between right eigenvectors can weaken the discontinuity of an occupied-state projector and change the coefficient of logarithmic entanglement scaling~\cite{XiaoRyu2026}. The present calculation does not use that many-body construction, and $\Sresp$ is not an entanglement entropy. The corresponding single-particle statement is that nonorthogonal right modes, together with their biorthogonal left partners, reshape the residues $A_\nu^{ab}(\omega)$ and therefore the interference terms in Eq.~\eqref{eq:interference-term}. The separation between $\Rcoh$ and $\Rmix$ consequently probes nonorthogonal modal geometry as well as spectral overlap.}

The structure of Eq.~\eqref{eq:rcoh} is closely related to a Kubo spectral decomposition. For a closed Hermitian system with eigenstates $|i\rangle$ and energies $\epsilon_i$, a generic response kernel contains terms
\begin{equation}
  \sum_{i,j}
  \frac{
  \langle i|A|j\rangle
  \langle j|B|i\rangle
  }
  {\omega-(\epsilon_j-\epsilon_i)+\ii 0^+}
  \left(
  p_i-p_j
  \right),
  \label{eq:kubo-generic}
\end{equation}
where $p_i$ are occupation weights. In electronic transport these weights may be Fermi factors. In the photonic setting considered here, they are instead determined by the drive, dissipation, and the chosen steady state. For a passive linear response around the vacuum, the one-particle Green function in Eq.~\eqref{eq:retarded-green} is the most direct object. Replacing a Hermitian Hamiltonian by $\Heff$ changes the ordinary matrix elements into biorthogonal residues, which is the origin of the distinction between Eqs.~\eqref{eq:rcoh} and \eqref{eq:rmix}.

The use of complex poles also makes the role of linewidths explicit. For a passive non-Hermitian photonic lattice we have $\Omega_\nu = \omega_\nu - \frac{\ii}{2}\kappa_\nu,$
so that $\frac{1}{\omega-\Omega_\nu}
  =  \frac{1}{\omega-\omega_\nu+\ii\kappa_\nu/2}$.
The linewidth is therefore intrinsic.
A small numerical broadening is still useful for plotting and for comparing different parameter values on the same frequency grid.

\section{Conclusion}
\rev{We have analyzed the linear optical response of a finite dissipative Hatano-Nelson chain using its biorthogonal Green function. 
The present treatment is intentionally restricted to the effective single-particle response.
The coherent and modal-weight channels separate interference among complex modal amplitudes from the distribution of their individual weights, while $\Neff$ measures how many poles contribute to a selected port-to-port signal. The spectral and time-domain calculations recover the expected boundary sensitivity: periodic boundaries support a complex spectral loop, whereas open boundaries produce directed propagation and right-edge accumulation. Loss increases modal overlap, while disorder reduces the response participation. In the finite $(g,W)$ map, the response-weighted center remains close to the skin boundary over moderate disorder even as $\Neff$ falls, showing that modal dephasing is detected before the spatial bias is removed. Disorder-ensemble averages confirm that this is a finite-size crossover rather than a thermodynamic transition. Taken together, the results identify the response participation number as a sensitive complement to spatial skin diagnostics in non-reciprocal photonic lattices.}



\clearpage
\begin{thebibliography}{99}

\bibitem{HatanoNelson1996}
Hatano, Naomichi, and David R. Nelson. "Localization transitions in non-Hermitian quantum mechanics." Physical review letters 77.3 (1996): 570.

\bibitem{HatanoNelson1997}
Hatano, Naomichi, and David R. Nelson. "Vortex pinning and non-Hermitian quantum mechanics." Physical Review B 56.14 (1997): 8651.

\bibitem{Kubo1957}
Kubo, Ryogo. "Statistical-mechanical theory of irreversible processes. I. General theory and simple applications to magnetic and conduction problems." Journal of the physical society of Japan 12.6 (1957): 570-586.

\bibitem{Carusotto2013}
Carusotto, Iacopo, and Cristiano Ciuti. "Quantum fluids of light." Reviews of Modern Physics 85.1 (2013): 299-366.

\bibitem{ElGanainy2018}
El-Ganainy, Ramy, et al. "Non-Hermitian physics and PT symmetry." Nature Physics 14.1 (2018): 11-19.

\bibitem{Miri2019}
Miri, Mohammad-Ali, and Andrea Alu. "Exceptional points in optics and photonics." Science 363.6422 (2019): eaar7709.

\bibitem{Parto2021}
Parto, Midya, et al. "Non-Hermitian and topological photonics: optics at an exceptional point." Nanophotonics 10.1 (2020): 403-423.

\bibitem{Lee2016}
Lee, Tony E. "Anomalous edge state in a non-Hermitian lattice." Physical review letters 116.13 (2016): 133903.

\bibitem{YaoWang2018}
Yao, Shunyu, and Zhong Wang. "Edge states and topological invariants of non-Hermitian systems." Physical review letters 121.8 (2018): 086803.

\bibitem{YaoSongWang2018}
Yao, Shunyu, Fei Song, and Zhong Wang. "Non-hermitian chern bands." Physical review letters 121.13 (2018): 136802.

\bibitem{Kunst2018}
Kunst, Flore K., et al. "Biorthogonal bulk-boundary correspondence in non-Hermitian systems." arXiv preprint arXiv:1805.06492 (2018).

\bibitem{YokomizoMurakami2019}
Yokomizo, Kazuki, and Shuichi Murakami. "Non-Bloch band theory of non-Hermitian systems." Physical review letters 123.6 (2019): 066404.

\bibitem{Brody2014}
Brody, Dorje C. "Biorthogonal quantum mechanics." Journal of Physics A: Mathematical and Theoretical 47.3 (2014): 035305.

\bibitem{Beenakker1997}
Beenakker, Carlo WJ. "Random-matrix theory of quantum transport." Reviews of modern physics 69.3 (1997): 731.

\bibitem{AkkermansMontambaux2007}
Akkermans, Eric, and Gilles Montambaux. Mesoscopic physics of electrons and photons. Vol. 1. Cambridge: Cambridge university press, 2007.

\bibitem{ForghieriParis2026}
Forghieri, Gaia, and Matteo GA Paris. "Engineering entanglement and transport in interacting quantum walks with tailored potentials." arXiv preprint arXiv:2606.17825 (2026).

\bibitem{AroucaLeeMoraisSmith2020}
Arouca, R., C. H. Lee, and C. Morais Smith. "Unconventional scaling at non-Hermitian critical points." Physical Review B 102.24 (2020): 245145.

\bibitem{HuWangLianWang2025}
Hu, Yu-Min, et al. "Many-body non-Hermitian skin effect with exact steady states in the dissipative quantum link model." Physical review letters 135.26 (2025): 260401.

\bibitem{GangulyAgarwalla2026}
Ganguly, Katha, and Bijay Kumar Agarwalla. "Full counting statistics for boundary driven transport in the presence of correlated gain and loss channels." Physical Review B 114.5 (2026): 055410.

\bibitem{XiaoRyu2026}
Xiao, Zhenyu, and Shinsei Ryu. "Anomalous entanglement scaling from eigenvector nonorthogonality in critical non-Hermitian free fermions." arXiv preprint arXiv:2607.25256 (2026).

\bibitem{LiZeng2026}
Li, Pu-Xuan, and Qi-Bo Zeng. "Tunable non-Hermitian skin effect and topological phases in ladders with staggered nonreciprocal inter-leg hopping." arXiv preprint arXiv:2607.11454 (2026).

\bibitem{GardinerZoller}
Gardiner, Crispin, and Peter Zoller. Quantum noise: a handbook of Markovian and non-Markovian quantum stochastic methods with applications to quantum optics. Springer Science \& Business Media, 2004.

\bibitem{Garcia2015}
García, Jose H., Lucian Covaci, and Tatiana G. Rappoport. "Real-space calculation of the conductivity tensor for disordered topological matter." Physical review letters 114.11 (2015): 116602.

\end{thebibliography}
\end{document}